\documentclass[aps,prl,twocolumn,superscriptaddress,a4paper,nofootinbib]{revtex4-2}
\usepackage[dvips]{graphicx}
\usepackage{color}
\usepackage{amsmath,amssymb,braket,bm,hyperref,tikz,tikz-feynman,mathbbol,xcolor}

\newcommand{\dd}{\mathrm{d}}
\newcommand{\hd}{\hat{\mathrm{d}}}
\newcommand{\hdelta}{\hat \delta}
\newcommand{\hop}{\mathbb{h}}
\newcommand{\helicity}{{\eta}}
\newcommand{\Smatrix}{{S}}
\def\SoftTO#1#2#3{{\mathcal{A}(#1, #2 | #3)}}
\def\SoftOT#1#2#3{{\mathcal{A}(#1 | #2, #3)}}

\renewcommand{\[}{\begin{equation}\begin{aligned}}
\renewcommand{\]}{\end{aligned}\end{equation}}

\renewcommand{\v}[1]{\bm{#1}}
\definecolor{allOrderBlue}{rgb}{0.4,0.5,1}

\definecolor{emerald}{rgb}{0.31, 0.78, 0.47}

\def\draftnote#1{{\textcolor{green}{\it #1}}}

\tikzset{rubout/.style={preaction={draw=white,line width=8pt}}}

\begin{document}

\title{Supertranslations from Scattering Amplitudes}

\author{Asaad Elkhidir}
\affiliation{Higgs Centre for Theoretical Physics, School of Physics and Astronomy, The University of Edinburgh, Edinburgh EH9 3JZ, Scotland, UK}
\author{Donal O'Connell}
\affiliation{Higgs Centre for Theoretical Physics, School of Physics and Astronomy, The University of Edinburgh, Edinburgh EH9 3JZ, Scotland, UK}
\author{Radu Roiban}
\affiliation{Institute for Gravitation and the Cosmos,
Pennsylvania State University,
University Park, PA 16802, USA\\
Institute for Computational and Data Sciences,
Pennsylvania State University,
University Park, PA 16802, USA}

\begin{abstract}
On-shell methods have found a new application to local observables such as asymptotic radiation fields and gravitational waveforms.
While these observables are invariant under small gauge transformations, they are known to depend on a choice of asymptotic gauge; in gravity on asymptotically Minkowski spacetimes, this is a choice of BMS frame.
In this letter, we provide a method for capturing these supertranslations, to all orders in perturbation theory, using the on-shell framework of scattering amplitudes.

\end{abstract}

\maketitle

\section{Introduction}

The literature on applications of scattering amplitudes to local asymptotic observables, such as the gravitational waveform, relies on a certain large-distance approximation~\cite{Cristofoli:2021vyo,DiVecchia:2022owy,Kosower:2022yvp,Herderschee:2023fxh,Brandhuber:2023hhy,Georgoudis:2023lgf,Elkhidir:2023dco,DeAngelis:2023lvf,Brandhuber:2023hhl,Aoude:2023dui,Georgoudis:2023eke,Bohnenblust:2023qmy,Fernandes:2024xqr,Bini:2024rsy,Georgoudis:2024pdz,Aoki:2024bpj,Brunello:2024ibk,Heissenberg:2024umh,Falkowski:2024bgb,Brandhuber:2024qdn,Fucito:2024wlg}.
Consider, the perturbative mediator operator (here graviton, but also photon or gluon):
\[
\label{eq:statPhase}
&\hop_{\mu\nu}(x) = \sum_\helicity \int \dd\Phi(k) \, e^\helicity_{\mu\nu}(k) a_\helicity(k) \, e^{-i k \cdot x} + \textrm{h.c.} \\
&\simeq \sum_\helicity\frac{-i}{4\pi |\v{x}|} \int_0^\infty  \hd \omega\, e^\helicity_{\mu\nu}(k) a_\helicity(k) e^{-i \omega u} |_{k = \omega n} + \textrm{h.c.} \,,
\]
where $u = x^0 - |\v x|$ is the retarded time and $n^\mu = (1, \hat{\v{x}})$.
The integration on the first line is over the on-shell phase space of a massless mediator, so that
\[
\dd\Phi(k) \equiv \hd^4 k \, \hdelta_+(k^2)\,, \quad  \hd k &\equiv \frac{\dd k}{2\pi} \,, \quad \hdelta(k) \equiv 2 \pi \, \delta(k) \\
\delta_+(k^2) &\equiv \delta(k^2) \Theta(k^0).
\]
The approximation leading to the second line of Eq.~\eqref{eq:statPhase} assumes that $\omega |\v{x}| \gg 1$ in the exponent on the first line.
%holds under the assumption that the quantity $\omega |\v{x}|$ in 
%the phase of the exponential function is large.

However, this assumption is certainly false in important situations even for very large $|\v{x}|$. 

Take, e.g. the gravitational waveform sourced when two point-like objects scatter.
Then the difference between the metric in the far future and the far past is non-zero, because the different initial and final velocities of the objects lead to different linearised Schwarzschild solutions. 
Correspondingly, the metric perturbation at fixed large distance $|\v x|$ will contain modes of frequency of order $1/|\v{x}|$ or smaller. 
A possible solution is to avoid the large-distance expansion and define creation operators at finite distance and time, as suggested in~\cite{Collins:2019ozc} in the development of a novel perspective on the LSZ reduction. 

In this letter we take a different approach and describe a method of capturing the contributions of these low-frequency modes in the spirit of the on-shell approach to scattering amplitudes.
As has been reemphasised~\cite{Caron-Huot:2023vxl} recently, a traditional way of incorporating the effects of low-frequency modes is to include the off-shell Coulombic field due to the initial particles.
Absence of gauge invariance for off-shell gravitons 
leads however to a complicated formalism.
%
%This approach has its disadvantages because taking 
%gravitons off-shell leads to a complicated formalism.
Here, following~\cite{Bini:2024rsy}, we will instead deform the on-shell condition of matter particles in a precise manner which will be explained below.
We will see that in electromagnetism, the Coulombic modes do not modify the expectation value of the gauge field beyond leading order.
In gravity, these modes lead to a BMS supertranslation of the gravitational strain, which can be computed to all orders in perturbation theory.

\section{Coherent background}

An appropriate quantum state for a single scalar point-like particle in the classical approximation is
\[
\label{eq:bareState}
\ket{\psi} = \int \dd\Phi(p) \, \varphi(p) \, \ket{p} \,,
\]
where $\varphi(p)$ is a momentum-space wavepacket which is steeply-peaked around some expected classical momentum.
Since there is no suppression of (classical) emission of low-frequency mediators, emission from a charged particle effectively 
creates its classical Coulombic field.
%if the particle is charged emission of such mediators effectively 
%creates the classical Coulombic field of the particle. 

Fig.~\ref{fig:resummation} captures schematically the low-frequency contributions to the final-state expectation value~\cite{Cristofoli:2021vyo} of the mediator operator in Eq.~\eqref{eq:statPhase}.
The right-hand side of the left diagram contains a $1\rightarrow 1+n$ amplitude, whose support is only on such modes.
Soft-limit properties of amplitudes and $\hbar$ counting then imply that only the classically-singular (superclassical) terms in this amplitude can contribute.
%, which can be confirmed though $\hbar$ counting. 
These are shown in the right diagram of Fig.~\ref{fig:resummation}.
As a means to enhance the large-distance approximation it is thus appropriate to resum their contributions. 

\iffalse
This can be understood by considering the final-state expectation value of the mediator field~\ref{eq:statPhase}, which gives its asymptotic value~\cite{Cristofoli:2021vyo}. Fig.~\ref{fig:resummation} captures schematically these low-frequency contributions; the right-hand side of the left diagram contains a $1\rightarrow 1+n$ amplitude, whose support is only on such modes. Soft-limit properties of amplitudes imply that in this limit only the classically-singular (superclassical) terms in this amplitude can contribute; they are shown in the right diagram of Fig.~\ref{fig:resummation}, and $\hbar$ counting indicates that they are classical.
%
As a means to enhance the large-distance approximation it is appropriate resum their contributions. 
\fi

\tikzset{
        graviton/.style={decorate,line width=0.1mm, decoration=
        {snake,amplitude=.5mm, segment length=2.0mm}
        },
        photon/.style={decorate, line width=0.4mm, decoration={snake} %, draw=red
        },
        etc/.style={decorate, dotted, decoration={} %, draw=red
        },
        massive/.style={postaction={decorate},
                line width=.5mm,
        },
        masslessWithArrow/.style={postaction={decorate},
                line width=0.4mm,
                decoration={
                        markings,
                         v}
        },
        massiveWithArrow/.style={postaction={decorate},
                line width=0.2mm,
                decoration={
                        markings,
                        mark=at position 0.5 with {\arrow{latex}}}
        },
        massiveWithTwoArrows/.style={postaction={decorate},
                line width=0.2mm,
                decoration={
                        markings,
                        mark=at position 0.25 with {\arrow{latex}}
                        markings,
                        mark=at position 0.95 with {\arrow{latex}}}
        },
        unitaryCut/.style={postaction={draw,densely dashed,blue,thin},
                line width = 0.15cm,white
        },
        cross/.default={1pt}
}
\begin{figure}
\centering
\begin{minipage}{0.23\textwidth}
\begin{tikzpicture}[scale=1.3]
    \draw[massiveWithTwoArrows] (-0.35, 0.9) -- (1, 0.9);
    \draw[massiveWithTwoArrows] (-0.35, -0.5) -- (1, -0.5);
    \draw[massiveWithTwoArrows] (2.20, 0.9) -- (1, 0.9);
    \draw[massiveWithArrow] (2.20, -0.5) -- (1, -0.5);
    \draw[graviton] (.45, 0) -- (1.65, 0);
    \draw[graviton] (.45, 0.25) -- (1.65, 0.25);
    \draw[graviton] (.45, 0.5) -- (1.65, 0.5);
    \draw[graviton] (.4, -0.25) -- (1, -0.25);
    \draw[etc] (0.75, 0.9) -- (0.75, .6);
    \draw[etc] (1.25, 0.9) -- (1.25, .6);
    \fill[allOrderBlue] (0.25, 0.20) ellipse (0.20 and 0.75);
    \fill[allOrderBlue] (1.65, 0.45) ellipse (0.15 and 0.5);
     \draw[unitaryCut] (1, 1.0) -- (1, -0.60);
    \node at (1.2,-0.25) {${K}$};
\end{tikzpicture}
\end{minipage}
\begin{minipage}{0.23\textwidth}
\begin{tikzpicture}[scale=1.3]
    \draw[massiveWithTwoArrows] (-0.35, 0.90) -- (1, 0.90);
    \draw[massiveWithTwoArrows] (-0.35, -0.5) -- (1, -0.5);
    \draw[massiveWithTwoArrows] (2.20, 0.9) -- (1, 0.9);
    \draw[massiveWithArrow] (2.20, -0.5) -- (1, -0.5);
    \draw[graviton] (.45, 0) -- (1, 0);
    \draw[graviton] (.45, 0.25) -- (1, 0.25);
    \draw[graviton] (.45, 0.5) -- (1, 0.5);
    \draw[graviton] (.40, -0.25) -- (1, -0.25);
%    \draw [graviton,in=180,out=-90] (1.5,1) to (1, 0.5);  
%    \draw [graviton,in=180,out=-90] (1.75,1) to (1, 0.25);
%    \draw [graviton,in=180,out=-90] (2,1) to (1, 0);
    \draw[graviton] (1.5,0.9) .. controls (1.35, 0.5) .. (1, 0.5);
    \draw[graviton] (1.75,0.9) .. controls (1.65, 0.35) .. (1, 0.25);
    \draw[graviton] (2,0.9) .. controls (1.9,0.15) .. (1, 0);
    \draw[etc] (0.75, 0.9) -- (0.75, .6);
    \draw[etc] (1.25, 0.9) -- (1.25, .6);
    \fill[allOrderBlue] (0.25, 0.20) ellipse (0.20 and 0.75);
 %   \fill[blue!50] (1.75, 0.5) ellipse (0.25 and 0.55);
     \draw[unitaryCut] (1, 1.0) -- (1, -0.60);
     \draw[unitaryCut] (1.36, 1.0) -- (1.36, 0.8);
     \draw[unitaryCut] (1.63, 1.0) -- (1.63, 0.8);
     \draw[unitaryCut] (1.88, 1.0) -- (1.88, 0.8);
    \node at (1.2,-0.25) {${K}$};
\end{tikzpicture}
\end{minipage}
\caption{Soft-mediator contribution to the asymptotic expression of the mediator field. The dashed vertical line crossing all horizontal lines represent the final state. }
\label{fig:resummation}
\end{figure}

A little work following Refs.~\cite{Monteiro:2020plf,Monteiro:2021ztt} shows that these $1\rightarrow 1+n$ amplitudes resum into an exponential factor~$U$
\[
\ket{\psi} \rightarrow U \ket{\psi} = \int \dd\Phi(p) \varphi(p) e^{i Q_s} \ket{p} \, ,
\label{eq:statedressing}
%\label{eq:Ustate}
\]
dressing the asymptotic state~\eqref{eq:bareState}.
The mechanism is very similar to that of Refs.~\cite{Monteiro:2020plf,Monteiro:2021ztt}, though now in Minkowski signature: only half-ladder diagrams contribute, 
%cf. Fig.~\ref{fig:resummation}, 
they all have equal numerators, all matter propagators are linear and the eikonal identity~\cite{Akhoury:2011kq} implies that all of them are on-shell. The factorial factor needed for exponentiation is the symmetry factor for the identical soft mediators in the final state. The exponent is 
%\begin{equation}
%\SoftOT{a}{b}{c} ~~~ \SoftTO{a}{b}{c}
%\end{equation}
\begin{align}
\label{eq:defOfSoftCharge}
\hspace{-5pt}
Q_s \equiv \sum_\helicity & \int \dd\Phi(k) \hdelta(2 p \cdot k)  %\mathcal{A}_3(p+k \rightarrow p, k_\helicity) 
\SoftOT{p+k}{p}{k_\helicity}
\, 
a^\dagger_\helicity(k)  
%\nonumber\\ &
+ \textrm{h.c.}\, ,
\end{align}
where the sum  is over the helicity $\helicity$ of the mediator and $\SoftOT{a}{b}{c}$ is the $(a\rightarrow b,c)$ one-mediator amplitude. There is one such factor for each matter particle of momentum $p$.
The delta functions and on-shell conditions imply that the integrand has support only for vanishing mediator frequency. When forming observables using the state in Eq.~\eqref{eq:statedressing} we will regularize the delta functions allowing low-frequency modes to possibly contribute.
For reasons which will become clear, we will also refer to $Q_s$ as the ``soft charge''.

\iffalse
The exponentiation mechanism is very similar to that of Refs.~\cite{Monteiro:2020plf,Monteiro:2021ztt}, though now in Minkowski signature: only half-ladder diagrams contribute, cf. Fig.~\ref{fig:resummation}, they all have equal numerators, all matter propagators are linear 
%in the incoming matter momentum 
and the eikonal identity~\cite{Akhoury:2011kq} implies that all of them are in fact on-shell. The factorial factor needed for exponentiation is the symmetry factor associated with the number of identical soft mediators in the final state.
\fi
\iffalse
The explicit sum in Eq.~\eqref{eq:defOfSoftCharge} is over the helicity $\helicity$ of the mediator; another sum over the matter particles is implicitly assumed.
The delta functions and on-shell conditions imply that the integrand has support only for vanishing mediator frequency. 
When forming observables using the state in Eq.~\eqref{eq:statedressing} we will see that infrared singularities require us to regulate the delta functions allowing low-frequency modes to contribute.
%
The operator $Q_s$ which appears here is a displacement operator, generating a coherent state of mediators.
For reasons which will become clear, we will also refer to $Q_s$ as the ``soft charge''.
\fi

The operator $Q_s$ in Eqs.~\eqref{eq:statedressing},\eqref{eq:defOfSoftCharge} is a displacement operator, generating a coherent state of mediators. 
To demonstrate its relation to the Coulombic background we recall 
that the gravitational three-point amplitude is
%To see that it indeed induces the Coulombic background, we recall 
%that the gravitational three-point amplitude is
\[
\label{eq:grMassiveA3}
%\mathcal{A}_3(p+k \rightarrow p, k_\helicity) 
\SoftOT{p+k}{p}{k_\helicity}= \kappa e^{\helicity *}_{\mu\nu} p^\mu p^\nu \,,
\]
where $\kappa^2 = 32 \pi G$, so that
\[
\label{eq:linSchw}
\kappa \braket{\psi | e^{-i Q_s} \hop^{\mu\nu}(x) e^{i Q_s} | \psi} = \frac{4G}{|\v x|} \frac{\Pi^{\mu\nu,\rho\sigma} p_\rho p_\sigma}{p \cdot n} \,.
\]
The physical state projector $\Pi = \sum_\helicity e_\helicity\otimes e^*_\helicity$ arises because of the identification of the helicity labels in the perturbative graviton operator and the soft charge due to the commutation relations of the graviton creation and annihilation operators.
Writing $p = m v$, we obtain the linearised metric $h_{\mu\nu}^\textrm{Schw}(x)$ of a Schwarzschild black hole of mass $m$ and velocity vector $v$, which is the same as the metric of a point particle of the same mass and velocity. 
Notice that we evaluated the metric operator in Eq.~\eqref{eq:linSchw} after using the approximation of Eq.~\eqref{eq:statPhase},
\[
\label{eq:displacement}
 e^{-i Q_s} \hop_{\mu\nu}(x) e^{i Q_s} = \hop_{\mu\nu}(x)  + h_{\mu\nu}^\textrm{Schw}(x) \, .
\]
In effect we are {\em defining} the metric operator by the use of this approximation as part of our prescription for incorporating Coulombic modes in an on-shell approach.
This is only possible because the field point $x$ is taken to be far from massive sources.

\iffalse
Note that we may express this result as
\[
\label{eq:displacement}
 e^{-i Q_s} \hop_{\mu\nu}(x) e^{i Q_s} = \hop_{\mu\nu}(x)  + h_{\mu\nu}^\textrm{Schw}(x) \,,
\]
where $h_{\mu\nu}^\textrm{Schw}(x)$ is the linearised Schwarzschild metric. 
\fi

\section{The hard charge}

We will soon be interested in evolving our initial state with the full $S$ matrix. As a step in this direction, let us examine the commutator $[S, Q_s]$ of the soft charge with the $S$ matrix. To that end, consider 
%the quantity
\begin{align}
\label{eq:defCommutator}
C \equiv \sum_\helicity \int \dd\Phi(k) \hdelta(2 p \cdot k) %\mathcal{A}_3(p+k \rightarrow p, k_\helicity) 
\SoftOT{p+k}{p}{k_\helicity}\, [\Smatrix, a^\dagger_\helicity(k)] \,.
\end{align}
Because of the small-frequency support, $k = \omega n$ and $\omega \rightarrow 0$, following from the on-shell conditions, we evaluate the commutator $[S, a^\dagger_\helicity(k)]$ in the soft limit.
We work to leading order in $\omega$, 
%\draftnote{Can $\omega^0$ matter?} 
restricting to situations in the classical approximation which involve scattering of some fixed number of stable and distinguishable massive particles. Self-consistently, we assume that the energies of the incoming particles are below the massive-particle pair-production threshold.
The scattering may nevertheless involve any number of incoming or outgoing mediators.

\iffalse
Because the on-shell condition for the massless mediator has support only at small frequencies, we must evaluate the commutator $[S, a^\dagger_\helicity(k)]$ in the soft limit $k = \omega n$ for $\omega \rightarrow 0$.
We work to leading order in soft $\omega$, 
\draftnote{Can $\omega^0$ matter?} 
restricting to situations in the classical approximation which involve scattering of some fixed number of stable and distinguishable massive particles.
Self-consistently, 
%because of the restriction to the classical approximation, 
%we assume that our particles are scattered at energies below 
%the pair-production threshold of massive particles.
we assume that the energies of the incoming particles are below 
the massive-particle pair-production threshold.
The scattering may nevertheless involve any number of incoming or outgoing mediators.
\fi

The commutator $[S, a^\dagger_\helicity(k)]$ is nonzero only for $S$ matrix elements involving an incoming soft mediator. Thus, we write $C = C_M + C_0$ where the two terms correspond to the mediator attaching to a massive and a massless line, respectively, and discuss the two terms separately.

\iffalse
For a given amplitude in the $S$ matrix, the commutator $[S, a^\dagger_\helicity(k)]$ vanishes unless the amplitude involves an incoming soft mediator.
Let us first consider the case where this soft mediator attaches to a massive particle.
That is, we write $C = C_M + C_0$ where $C_0$ involves the mediator attaching to massless lines, and will be discussed below.
Because of our assumptions, the massive particle in $C_M$ is present in both the initial and final state, and therefore there are two related amplitudes of interest shown in figure~\ref{fig:massiveCase}.
As we will see, the contributions from these two diagrams sum up into a simple result for the commutator.
\fi

\begin{figure}
\centering
\begin{minipage}{0.23\textwidth}
\begin{tikzpicture}[scale=1.0] 
\begin{feynman}
  \vertex (vMid) {};
  \vertex [above left=0.6 and 1.8 of vMid] (ul) {$P$};
  \vertex [above right=0.6 and 1.8 of vMid] (ur) {$P'$};
  \vertex [below left=0.6 and 1.8 of vMid] (dl) {};
  \vertex [below right=0.6 and 1.8 of vMid] (dr) {};
  \vertex [above=0.2 of vMid] (midg1);
  \vertex [left=1.4 of midg1] (ig1);
  \vertex [right=1.4 of midg1] (og1);
  \vertex [below=0.2 of vMid] (midg2);
  \vertex [left=1.4 of midg2] (ig2);
  \vertex [right=1.4 of midg2] (og2);
  \vertex [right=1.4 of vMid] (og3);
  \draw [thick] (ul) -- (ur);
  \draw [thick] (dl) -- (dr);
  \vertex [right=1 of ul] (vgamma);
  \vertex [above=0.7 of vgamma] (igamma);
  \vertex [above=0.7 of ul] (uul) {$p$};
  \vertex [above=0.7 of ur] (uur) {};
  \draw [thick] (uul) -- (uur);
  \path (igamma) to node [pos=0.5] {$\;\;\;\;\;k$} (vgamma);
  \diagram* {
    (igamma) -- [photon, red] (vgamma);
    (ig1) -- [photon] (og1);
    (ig2) -- [photon] (og2);
    (vMid) -- [photon] (og3);
  };
  \fill [allOrderBlue] (vMid) ellipse (8pt and 20pt);
\end{feynman}
\end{tikzpicture} 
\end{minipage}
\begin{minipage}{0.23\textwidth}
\begin{tikzpicture}[scale=1.0] 
\begin{feynman}
  \vertex (vMid) {};
  \vertex [above left=0.6 and 1.8 of vMid] (ul) {$P$};
  \vertex [above right=0.6 and 1.8 of vMid] (ur) {$P'$};
  \vertex [below left=0.6 and 1.8 of vMid] (dl) {};
  \vertex [below right=0.6 and 1.8 of vMid] (dr) {};
  \vertex [above=0.2 of vMid] (midg1);
  \vertex [left=1.4 of midg1] (ig1);
  \vertex [right=1.4 of midg1] (og1);
  \vertex [below=0.2 of vMid] (midg2);
  \vertex [left=1.4 of midg2] (ig2);
  \vertex [right=1.4 of midg2] (og2);
  \vertex [right=1.4 of vMid] (og3);
  \draw [thick] (ul) -- (ur);
  \draw [thick] (dl) -- (dr);
  \vertex [left=1 of ur] (vgamma);
  \vertex [above=0.7 of vgamma] (igamma);
  \vertex [above=0.7 of ul] (uul) {$p$};
  \vertex [above=0.7 of ur] (uur) {};
  \draw [thick] (uul) -- (uur);
  \path (igamma) to node [pos=0.5] {$\;\;\;\;\;k$} (vgamma);
  \diagram* {
    (igamma) -- [photon, red] (vgamma);
    (ig1) -- [photon] (og1);
    (ig2) -- [photon] (og2);
    (vMid) -- [photon] (og3);
  };
  \fill[allOrderBlue] (vMid) ellipse (8pt and 20pt);
\end{feynman}
\end{tikzpicture} 
\end{minipage}
\caption{A soft mediator (red) arising from dressing a charged particle of momentum $p$ may attach to another particle of momentum $P$, leading to two Feynman diagrams of this general type. Several massive particles could be present, and, in principle, any number of incoming or outgoing mediators. Note that the case $p = P$ may arise.}
\label{fig:massiveCase}
\end{figure}

Focusing first on $C_M$, our assumptions imply that the massive particle is present in both the initial and final state, and therefore there are two related amplitudes of interest, shown in Fig.~\ref{fig:massiveCase}. It is convenient to discuss them together.
Because of the soft limit, the $S$ matrix element associated with the left diagram in Fig.~\ref{fig:massiveCase} is, to all orders in the coupling,
\[
-\sum_\helicity  \int \dd\Phi(k, P, P') \, (i H(P, P'))\, 
\frac{
%\mathcal{A}_3(P, k_\helicity  \rightarrow P+k)
\SoftTO{P}{k_\helicity}{P+k}}{2 P\cdot k + i\varepsilon}  a_\helicity (k) \,,
\]
where $H(P,P')$ refers to the ``hard'' matrix element with all external particles other than the soft mediator, and we have explicitly written the annihilation operator of the soft particle.
Notice that ${\cal A}_3$ appearing here is more precisely a soft factor; however, at the leading order in $\omega$ the difference is negligible. 
Similarly, to this order we have neglected the difference between $H(P+k,P')$ and $H(P,P')$. 
In the same way, the $S$-matrix element associated with the right diagram is
\[
\sum_\helicity  \int \dd\Phi(k, P, P') \, (i H(P, P'))\, \frac{
%\mathcal{A}_3(P'-k, k_\helicity  \rightarrow P')
\SoftTO{P'-k}{k_\helicity}{P'}}{2 P'\cdot k - i\varepsilon}  a_\helicity (k) \,.
\]

Thus, the two contributions to $C_M$ contain the sums
\[\label{eq:XandXprime}
X &= \sum_\helicity  
%\mathcal{A}_3(P, k_\helicity  \rightarrow P+k)  
\SoftTO{P}{k_\helicity}{P+k}  
%\mathcal{A}_3(p + k \rightarrow p, k_\helicity )  
\SoftOT{p + k}{p}{k_\helicity}
\, ,
\\
X'&= \sum_\helicity  
%\mathcal{A}_3(P'-k, k_\helicity  \rightarrow P') 
\SoftTO{P'-k}{k_\helicity}{P'}  
%\mathcal{A}_3(p +k \rightarrow p, k_\helicity)
\SoftOT{p + k}{p}{k_\helicity}
\, ,
\]
which are the same to leading order in the $\hbar$ expansion but differ at the next-to-leading order. 
Their common leading-order value contributes to
Eq.~\eqref{eq:defCommutator} as 
%\Donal{Donal to remove irrelevant delta plus}
\[\label{eq:Cmassive}
C_M = \int \dd\Phi(k,P',P)\,\hdelta(2p\cdot k) X \, (i H(P,P'))& \\
\times  \left( \frac{1}{2P'\cdot k - i \varepsilon} - \frac{1}{2P\cdot k + i \varepsilon} \right)& \,.
\]
To obtain a classical result we must expand to one subdominant order in $\hbar$. Using momentum conservation for the
%that $P' = P + Q + k$ for the momentum 
configuration in Fig.~\ref{fig:massiveCase}, and that both $Q$ and $k$ are ${\cal O}(\hbar)$ the dominant 
%(classically-singular) 
term is proportional to 
%the integral
\[\label{eq:VanishingIntSC}
\int \dd\Phi(k)\,\hdelta(2p\cdot k) \hdelta(2 P\cdot k) \, .
\]
This integral vanishes~\cite{Bini:2024rsy}. 
%
%To see this we first note that it has triangle topology with three 
%cut lines. Similar integrals, with at least one cut matter leg 
%
Similar integrals with fewer cut matter lines
also occur at the next order in the $\hbar$ expansion and in the calculation of $C_0$. We will uniformly regulate them by using their scaling properties to set to unity the time-like component of the external momentum $P$ not entering the cut condition and shifting the one that does, $p$, as $p\rightarrow p + m \beta P/P^0$ in the quadratic form $(p+k)^2 = m^2$ of the cut condition. $\beta$ is a regulator to be set to zero at the the end. The deformed condition is
\[ 
2 p\cdot k + 2 m \beta  p\cdot P /P^0  = 0\, .
\]
Terms proportional to $P \cdot k$ are dropped because they either vanish or yield scaleless integrals. Terms ${\cal O}(\beta^2)$ are dropped because $\beta\rightarrow 0$.
With this regularization $\hdelta(2P\cdot k)$ in Eq.~\eqref{eq:VanishingIntSC} has no support so the integral 
vanishes. 

The classical contribution due to the difference between $X$ and $X'$ in Eq.~\eqref{eq:XandXprime} is proportional to 
\[
\label{eq:vanishingintegral}
I(p, P) = \int \dd\Phi(k)\,\frac{ \hdelta(2p\cdot k)  }{2 P\cdot k -i \varepsilon}  \, ,
\]
while that of Eq.~\eqref{eq:VanishingIntSC} is proportional to
\[
\label{eq:vanishingintegral}
\int \dd\Phi(k)\, \frac{Q \cdot k \, \hdelta(2p\cdot k)}{(2 P\cdot k -i \varepsilon)^2} = -\frac{1}{2}Q\cdot \frac{\partial}{\partial P} I(p, P)\,.
\]
It is not difficult to see that ($Z=\sqrt{(p\cdot P)^2-p^2P^2}$)
\[
\label{eq:vanishingintegral}
I(p, P) = -\frac{\pi}{2Z}
\ln \frac{p\cdot P + Z }{p\cdot P - Z}
\,.
\]
The case $p=P$, corresponding to a zero-energy graviton starting and ending on the same matter line, can either be obtained as the limit $p\rightarrow P$ of $I(p, P)$ or independently. Care is needed in the latter approach to account for a singular propagator regularized by the $\beta$ deformation.  

Thus, the matter contribution $C_M$ to the commutator $[S, Q_s]$ can be written as
%is on the one hand nonvanishing and on the other 
%that it can be written as 
\[
\label{eq:Qhardmatter}
C_M = -[S, Q_{h, M}]
\qquad Q_{h, M} = \sum_i Q_{h, i} a^\dagger_i a_i \ ,
\]
where the sum is over all the matter particles and $Q_{h, i}$, which is a linear combination of $I(p, P)$ and its derivatives, is itself a sum over all the matter particles. 

In electromagnetism, mediators only interact by attaching to external massive (charged) particles, and thus $[\Smatrix, Q_s+Q_{h, M}] = 0$.
%\[\relax
%[\Smatrix, Q_s] = 0\,.
%\]
%In other words, the displacement operator which generates the 
%Coulomb mode commutes with the $S$ matrix in electromagnetism. 
We will discuss below consequences of this relation.

\begin{figure}
\centering
\begin{tikzpicture}[scale=1] 
\begin{feynman}
  \vertex (vMid) {};
  \vertex [above left=0.6 and 1.8 of vMid] (ul) {$P$};
  \vertex [above right=0.6 and 1.8 of vMid] (ur) {$P'$};
  \vertex [below left=0.6 and 1.8 of vMid] (dl) {};
  \vertex [below right=0.6 and 1.8 of vMid] (dr) {};
  \vertex [above=0.2 of vMid] (midg1);
  \vertex [left=1.8 of vMid] (vg1) {$K$};
  \vertex [right=1.4 of midg1] (og1);
  \vertex [below=0.2 of vMid] (midg2);
  \vertex [right=1.4 of midg2] (og2);
  \vertex [right=1.4 of vMid] (og3);
  \draw [thick] (ul) -- (ur);
  \draw [thick] (dl) -- (dr);
  \vertex [left=1.0 of vMid] (vgamma);
  \vertex [above=1.3 of vgamma] (igamma);
  \vertex [above=0.7 of ul] (uul) {$p$};
  \vertex [above=0.7 of ur] (uur) {};
  \diagram* {
    (igamma) -- [photon, red, rubout] (vgamma);
    (midg1) -- [photon] (og1);
    (midg2) -- [photon] (og2);
    (vg1) -- [photon] (og3);
  };
  \draw [thick] (uul) -- (uur);
  \path (vgamma) to node [pos=0.75] {$\;\;\;\;k$} (igamma);
  \filldraw [fill=red] (vgamma) circle (2pt);
  \fill[allOrderBlue] (vMid) ellipse (8pt and 20pt);
\end{feynman}
\end{tikzpicture} 

\caption{In gravity and Yang-Mills theory, a soft mediator (red) may attach to external mediators.}
\label{fig:masslessCase}
\end{figure}

In Yang-Mills theory and in gravity soft modes may 
also interact with external mediators (gauge boson or gravitons).
Therefore we must consider the additional contributions to $C$ of the 
type shown in Fig.~\ref{fig:masslessCase}.
Unlike massive particles, there is no requirement for massless mediators to appear both in initial and final states. Moreover, since $Q_s$ is linear in $a^\dagger_\helicity$, we can focus on one external mediator at a time, bearing in mind that we must sum over {\em all} external gravitons {\em and all} external matter particles.
We find, in gravity,
\[\label{eq:MasslessIntegral}
C_0 = - \int \dd\Phi(K) i H(K) \int \dd\Phi(k) \hdelta(2 p \cdot k) \frac{\kappa^2 (K \cdot p)^2}{2 K \cdot k + i \varepsilon} \,.
\]
The soft integral does not vanish after the $\beta$ deformation~\cite{Bini:2024rsy} $p\rightarrow p + m \beta K/K^0$, and has both finite and divergent terms.
Denoting the latter by $Y(\beta, \epsilon)$ where $\epsilon=(4-d)/2$ is the dimensional regulator, and making use of its momentum (in)dependence, the result for the commutator of the $S$ matrix and the soft charge is
\[
C_0 &= 2  G \int \dd\Phi(K) \, (i H(K)) \, p \cdot K \Big[\log \left(\frac{p\cdot K}{m K^0}\right)+Y\Big] \\
&=- [S, Q_{h, 0}] \, .
\]
On the second line we defined the hard charge $Q_{h, 0}$ 
\begin{align}
\label{eq:hardChargeDef}
Q_{h, 0} = -&2G \, \sum_\helicity  \int \dd \Phi(K) \, p \cdot K \\
&\times \Big[
\log  \left(\frac{p\cdot K}{m K^0} \right)+Y(\beta, \epsilon)\Big] \, a^\dagger_\helicity (K) a_\helicity (K) \, ,
\nonumber
\end{align}
so that its commutator with the $S$ matrix cancels that of the soft charge. A sum over matter particles is implied.
It follows that the sum of the hard, $Q_h=Q_{h, 0}+Q_{h, M}$, and soft charges commutes with the $S$ matrix in gravity:
\[\relax \label{eq:commutes}
[\Smatrix, Q_s + Q_h] = 0 \,.
\]
It is also not difficult to see that the hard and soft charges also  commute with each other, $[Q_s, Q_h] = 0$.

Eq.~\eqref{eq:commutes} can be interpreted as an equivalence between inclusion of a vanishingly-soft mediator in an $\Smatrix$-matrix element and a particular rescaling of the $\Smatrix$-matrix element without that low frequency  particle.
In the remainder of this letter, we will discuss consequences for observables.

\iffalse
From the definitions of $Q_s$ (Eq.~\eqref{eq:defOfSoftCharge}) and $Q_h$ (Eq.~\eqref{eq:hardChargeDef}), an interpretation of this statement is that a low frequency mediator in the final state is equivalent to a particular rescaling of the $S$-matrix element without that low frequency  particle. 
%
In the remainder of this letter, we will discuss consequences for observables.
\fi
%%%%%
\iffalse
One may also establish, using the explicit expressions of the hard and soft charges, that they commute:
\[\relax
[Q_s, Q_h] = 0 \,.
\]
In the remainder of this letter, we will discuss consequences for observables.
\fi

\section{Supertranslations}

Consider the expectation value of the metric operator in the future asymptotic state $|\psi\rangle_\text{out} = \Smatrix \exp(i Q_s) |\psi\rangle$: 
\begin{align} \label{eq:introHardCharge}
h_{\mu\nu}(x) &\equiv\braket{\psi | e^{-i Q_s} \Smatrix^\dagger \hop_{\mu\nu}(x) \Smatrix e^{i Q_s} | \psi} \\
&\quad = 
\braket{\psi | e^{-i (Q_s + Q_h)} \Smatrix^\dagger \hop_{\mu\nu}(x) \Smatrix e^{i (Q_s + Q_h)} | \psi} \,. 
\nonumber
\end{align}
We introduced the hard charge above, noting that $Q_h \ket{\psi} = 0$. 
More generally, if the initial state includes a gravitational background, then inclusion of $Q_h$ in this way introduces a phase which cancels in the expectation value because it is a function only of incoming kinematic invariants; matter hard charges also cancel in the expectation value for a similar reason.
We have explicitly verified that $Q_{h, M}$ gives vanishing contributions to $\langle \hop_{\mu\nu}\rangle$ in processes with two incoming and two outgoing matter particles.
Since $[\Smatrix, Q_s+Q_h]=0$ and $[Q_s, Q_h]=0$, we may equivalently write,
using also Eq.~\eqref{eq:displacement},
%Since the total charge $Q_s + Q_h$ commutes with the $S$ matrix, 
%we may equivalently write
\[
h_{\mu\nu}(x) &= \braket{\psi | \Smatrix^\dagger e^{-i Q_h} e^{-i Q_s} \hop_{\mu\nu}(x) e^{i Q_s} e^{i Q_h} \Smatrix |\psi} \\
&= \braket{\psi | \Smatrix^\dagger e^{-i Q_h} \hop_{\mu\nu}(x) e^{i Q_h} \Smatrix |\psi} + h_{\mu\nu}^\textrm{Schw}(x) \,,
\label{eq:metricexpvalue}
\]
%using Eq.~\eqref{eq:displacement}.
Thus, to all orders, the effect of the soft charge is to introduce the linearised Schwarzschild term in the metric.
%expectation.

The effect of the hard charge can also be made explicit in Eq.~\eqref{eq:metricexpvalue} to all orders. Indeed, using Eq.~\eqref{eq:hardChargeDef}, it is not difficult to check that
\[\label{eq:HardChargePhase}
e^{-i Q_{h,0}} a_\helicity (k) e^{i Q_{h,0}} = e^{i \omega T(n)} a_\helicity (k) \,,
\]
where for each matter particle we have defined $T(n) $ as
\[
\label{eq:Tofn}
T(n) \equiv 2G (p \cdot n) \left( \log (v \cdot n) + Y(\beta, \epsilon)  \right)\,.
\]
%for each matter particle.
Here, we set $\omega = k^0$ and $n = k /\omega$ and made use of the relation between the momentum $p$ and the velocity $v$.
Notice that the ratio $K^\mu / K^0$, which appeared in the definition, Eq.~\eqref{eq:hardChargeDef}, of the hard charge, has been replaced by $n^\mu$ due to the action of the number operator on $a_\helicity(k)$. 

In view of the $u$ dependence of Eq.~\eqref{eq:statPhase}, the net result is a shift for each matter particle,
\[
\label{eq:VV}
u \rightarrow u + 2Gm \, v\cdot n \log v\cdot n \,,
\]
where we absorbed the regulator-dependent $Y(\beta, \epsilon)$ by an appropriate choice of the origin of retarded time.
As $n = (1, \hat{\v x})$ after use of Eq.~\eqref{eq:statPhase}, the shift in Eq.~\eqref{eq:VV} is an angle-dependent translation of the retarded time, i.e. a BMS supertranslation, which is in fact precisely that of Veneziano and Vilkovisky.
%
%Supertranslations, by definition, are angle-dependent translations 
%of this type, and in fact the shift in Eq.~\eqref{eq:VV} is precisely the Veneziano-Vilkovisky case.

Classically, the Veneziano-Vilkovisky supertranslation~\cite{Veneziano:2022zwh} relates the ``intrinsic'' and ``canonical'' BMS frames.
In the former~\cite{Veneziano:2022zwh}, the metric strain at ${\cal O}(G)$ is that of linearised Schwarzschild while in the latter the metric strain is zero~\cite{Flanagan:2015pxa}.
Our result demonstrates that the hard charge generates a large change of coordinates. Thus, while rooted in perturbative (quantum) field theory, our approach makes direct contact with the classical physics of large gauge transformations.

Eq.~\eqref{eq:VV} also shows that the Veneziano-Vilkovisky transformation connects the intrinsic and canonical frames to all orders in the coupling. This situation is reminiscent of the Kerr-Schild coordinates, in which all higher-order corrections to the linearised metric are absorbed by a judicious change of coordinates. 
Although the phase \eqref{eq:HardChargePhase} amounts to a simple coordinate shift in the waveform, it is essential, as noted in \cite{Georgoudis:2023eke} and further confirmed in \cite{Georgoudis:2024pdz,Bini:2024rsy}, to resolving differences between QFT~\cite{Brandhuber:2023hhy,Herderschee:2023fxh,Georgoudis:2023lgf,Elkhidir:2023dco} and Multipolar post-Minkowskian~\cite{Bini:2023fiz} results for the one-loop waveform.
%

\iffalse
Although the phase \eqref{eq:HardChargePhase} amounts to a simple coordinate shift in the waveform, it is crucial to take into account when comparing amplitudes-based approaches with the multipolar post-Minkowskian results of Ref.~\cite{Bini:2023fiz}. As noted in \cite{Georgoudis:2023eke}, and further confirmed in \cite{Georgoudis:2024pdz,Bini:2024rsy}, this phase contributes to resolving  differences between QFT~\cite{Brandhuber:2023hhy,Herderschee:2023fxh,Georgoudis:2023lgf,Elkhidir:2023dco} and classical GR~\cite{Bini:2023fiz} results for the one-loop waveform.
Our approach is rooted in (quantum) field theory, yet makes direct contact with the classical physics of large gauge transformations.
\fi

As another perspective, notice that the supertranslation of Eq.~\eqref{eq:VV} could be introduced at the level of correlation functions by supertranslating the quantum fields.
This 
%is consistent with 
echoes the suggestion of Ref.~\cite{Cristofoli:2022phh} that large gauge effects can be connected to a modified LSZ prescription.

\section{Angular Momentum}
It is well known that the radiated angular momentum suffers from a BMS ambiguity related to the contribution of the static field~\cite{Penrose:1964ge,Newman:1966ub,Ashtekar:1979iaf,Ashtekar:2019rpv, DiVecchia:2022owy,Compere:2019gft,Chen:2021kug,Chen:2021szm,Javadinezhad:2022ldc,Javadinezhad:2022hhl,Mao:2023evc,Riva:2023xxm,Javadinezhad:2023mtp,Mao:2024ryq}. 
%In recent years 
This issue has been sharpened~\cite{Damour:2020tta,Manohar:2022dea,Veneziano:2022zwh,Heissenberg:2024umh,DiVecchia:2022piu} in the context of the programme of computing radiative observables from scattering amplitudes. 
Static contributions in the intrinsic BMS frame
%, for example in the intrinsic BMS frame, 
lead to a perturbative enhancement of the angular momentum flux relative to the power counting of radiative 
%(five-point) 
amplitudes.
%On the other hand there is no contribution from static fields to the 
%angular momentum flux in the canonical BMS frame~\cite{Bonga:2018gzr} 
%since the shear is zero.
No such contributions occur in the canonical BMS frame~\cite{Bonga:2018gzr} since the shear is zero, consistent with the expected power counting from amplitudes.
%This is consistent with the expected power counting from amplitudes.

Thus, the leading contribution of static fields, if present, enters the radiated angular momentum at $\mathcal{O}(G^2)$.  In \cite{DiVecchia:2022owy}, this contribution was recovered by resumming the contribution of zero-frequency mediators into an eikonal operator\footnote{This is analogous to reassigning the rephasing in Eq.~\eqref{eq:statedressing} to the $S$ matrix rather than to the external states.}, which is then used to evaluate expectation value of the angular momentum operator. 

Here we discuss the contribution of these static fields using our approach to all orders in $G$.
We will recover the order $G^2$ contribution in the intrinsic frame, and we will show that the hard charge also contributes to the orbital angular momentum flux $L_{\alpha \beta}$ at $\mathcal{O}(G^4)$. 
%In order 
To work in the intrinsic frame, we start by evaluating the expectation value of the angular momentum operator in the future dressed particle state $|\psi\rangle_\text{out}$:
\[\label{eq:OrbitalExp}
L_{\alpha \beta} \equiv \bra{\psi} e^{- i Q_s} \Smatrix^\dagger \mathbb{L}_{\alpha \beta} \Smatrix e^{iQ_s}\ket{\psi},
\]
where
\[
\mathbb{L}_{\alpha \beta} = -  i \sum_\helicity  \int \dd \Phi(k) \; a_\helicity ^{\dagger}(k) \; k_{[\alpha} \stackrel{\leftrightarrow}{\partial}_{\beta]} a_\helicity (k) 
\]
is the orbital angular momentum operator and 
\[
f  \stackrel{\leftrightarrow}{\partial_{\mu}} g\equiv \frac{1}{2}\left( f \frac{\partial g}{\partial k^\mu} -  \frac{\partial f}{\partial k^\mu} g \right). 
\]
We may introduce the hard charge $Q_h$ as in Eq.~\eqref{eq:introHardCharge} and use Eq.~\eqref{eq:commutes}
%, as in Eq.~\eqref{eq:metricexpvalue}, 
%use the commutator~\eqref{eq:commutes} to 
to rewrite the orbital angular momentum as
\begin{align}
L_{\alpha \beta} =& -i  \sum_{ \helicity } \int \dd \Phi(k) \bra{\psi} \Smatrix^\dagger  e^{-i (Q_s + Q_h)}a_\helicity^{\dagger}(k) e^{i (Q_s +Q_h)} \nonumber\\
&\times  k_{[\alpha} \stackrel{\leftrightarrow}{\partial}_{\beta]} e^{-i (Q_s +Q_h)}a_\helicity(k) e^{i (Q_s + Q_h)} \Smatrix \ket{\psi} \ .
\label{eq:AMintegrand}
\end{align}
%%%%%%%%%%%%%%%%%%%%%%%%%%%%%%%%%%
%%% merged 2 equations L = and I = 
\iffalse
\[
L_{\alpha \beta} = -i  \sum_{ \helicity } \int \dd \Phi(k) I^\helicity_{\alpha \beta} \,.
\]
where $ I^\helicity_{\alpha \beta}$ is explicitly
\[\label{eq:AMintegrand}
 I^\helicity_{\alpha \beta} &= \bra{\psi} \Smatrix^\dagger  e^{-i (Q_s + Q_h)}a_\helicity^{\dagger}(k) e^{i (Q_s +Q_h)}  \\
&\times  k_{[\alpha} \stackrel{\leftrightarrow}{\partial}_{\beta]} e^{-i (Q_s +Q_h)}a_\helicity(k) e^{i (Q_s + Q_h)} \Smatrix \ket{\psi}.
\]
\fi
%%%%%%%%%%%%%%%%%%%%%%%%%%%%%%%%%%
This form allows us to exploit the relations
\[
& e^{-i(Q_s +Q_{h,0})} a^\dagger_\helicity (k) e^{i (Q_s +Q_{h,0})} = e^{-i \omega T({n})}  a^\dagger_\helicity(k) + \alpha^*_\helicity(k) ,\\
& e^{-i (Q_s +Q_{h,0})} a_\helicity(k) e^{i(Q_s + Q_{h,0})} = e^{i \omega T({n})}  a_\helicity(k) + \alpha_\helicity(k),
\]
where we define
\[\label{eq:alpha}
\alpha_\helicity(k) \equiv \hdelta(2 p \cdot k)  
\SoftOT{p + k}{p}{k_\helicity}
%\mathcal{A}_3(p -k \rightarrow p, k_\helicity) 
\,.
\]
Inserting this in \eqref{eq:AMintegrand}, there are three classes of term: those with two $\alpha$'s, those with one $\alpha$, and terms with no $\alpha$.
The terms with two powers of $\alpha$ vanish after integration over $k$ because the observable angular momentum tensor must be antisymmetric in $\alpha$ and $\beta$ but can only be built from the velocity vector (and the metric).

The linear-in-$\alpha$ terms contain the $\mathcal{O}(G^2)$ contribution to the angular momentum flux noted in \cite{Damour:2020tta,Manohar:2022dea,DiVecchia:2022owy}.  
This is because $\alpha$ is ${\cal O}(\kappa)$, and the rest of the matrix element in Eq.~\eqref{eq:AMintegrand} is ${\cal O}(\kappa^3)$ (or higher).
Note that the phase $i \omega T(n)$ drops out of these linear-in-$\alpha$ terms due to the delta function in Eq.~\eqref{eq:alpha}.
We conclude that the hard charge does not contribute to these terms.

Turning to the $\alpha$-independent contributions it is easy to see that the hard charge yields
%, through the phase $i \omega T$, 
an additional contribution to the orbital angular momentum flux since
\[
e^{- i \omega T(n)}k_{[\alpha} \frac{\stackrel{\leftrightarrow}{\partial}}{\partial k^{\beta]}} e^{i \omega T(n)} 
= 2iG k_{[ \alpha} p_{\beta]} \log (v \cdot n )\,. 
\]
%Overall, we find that the hard charge makes a contribution 
%$L^{\text{hard}}_{\alpha \beta}$ to the orbital angular 
%momentum flux given by
%
Overall, we find that the hard charge 
%makes a contribution $L^{\text{hard}}_{\alpha \beta}$ 
contributes to the orbital angular 
momentum flux as
%given by
\begin{align}
L^{\text{hard}}_{\alpha \beta}= & 2  G \sum_{X,h} \int \dd \Phi(k,\tilde{p}_1 ,\tilde{p}_2 , X) \bra{\psi} \Smatrix^\dagger  a_\helicity^{\dagger}(k) \ket{\tilde{p}_1 ,\tilde{p}_2 , X} \nonumber\\
&\times \bra{\tilde{p}_1 ,\tilde{p}_2 , X}a_\helicity(k)  \Smatrix \ket{\psi} k_{[ \alpha} p_{\beta]} \log (v \cdot n ),
\end{align}
where we have inserted a complete set of intermediate states. Noting that each of the matrix elements above contributes at $\mathcal{O}(\kappa^3)$, we see that the leading contribution of $L^\text{hard}_{\mu \nu}$ enters at $\mathcal{O}(G^4)$. Indeed, this contribution is known to arise from static fields \cite{Heissenberg:2024umh} and is associated with the nonlinear memory effect \cite{Christodoulou:1991cr, Wiseman:1991ss,Thorne:1992sdb}.

\section{Connection to Classical Treatment}

The classical charges are expectation values of the quantum operators. 
To make the connection with the classical BMS transformation more transparent, we will reorganize the frequency integrals in Eqs.~\eqref{eq:defOfSoftCharge} and \eqref{eq:hardChargeDef} as integrals over the retarded time, and also expose the parameter $T(n)$  of the transformation.

Focusing on the most familiar case of gravity and using the amplitude~\eqref{eq:grMassiveA3}, the soft charge~\eqref{eq:defOfSoftCharge} operator is
\[
Q_s = \sum_\helicity  \frac{1}{4} \int \hd \omega \, \hd^2 \Omega\, \omega \, \hdelta_+(\omega) \frac{ \kappa \; e_\helicity ^{\mu \nu} p_\mu p_\nu }{ n \cdot p}  a_\helicity (\omega n)
+ \textrm{h.c.} 
\]
where, as before, $n=k/\omega= (1, \v n(\theta, \phi))$ is a general null vector. We denote to the polar angles $\theta$ and $\phi$ as $n^A$.
\iffalse
We made a change of variables $k = \omega n$ where, for now, the vector $n^\mu = (1, \v n(\theta, \phi))$ is a general null vector.
We further write the polar angles $\theta$ and $\phi$ as $n^A$.
\fi

It is useful to choose a basis of polarisation vectors\footnote{We work in a mostly-minus signature.}
\[
\epsilon_A^\mu = \partial_A n^\mu = D_A n^\mu 
\quad A=1,2 \,,
\]
where $D_A$ is the covariant derivative on the (unit) sphere.
%which will be useful below.
These vectors have the property that
\[
\epsilon_A \cdot \epsilon_B = - \Omega_{AB} \,,
\]
where $\Omega_{AB}$ is the unit metric on the two-sphere.
We may use this basis to write the graviton polarisation tensor as
\[
e^\helicity_{\mu\nu} = e^\helicity_{AB} \epsilon^A_\mu \epsilon^B_\nu \,.
\]
(The $A, B$ indices are raised and lowered using $\Omega$.)
Tracelessness of $e^\helicity_{\mu\nu}$ implies that $e^\helicity_{AB} \Omega^{AB} = 0$.
The soft charge then becomes 
\[\label{eq:SoftCharge1}
Q_s =  \frac{1}{4}  & \int \hd^2 \Omega\,  \kappa \, \frac{ D_A (n \cdot p) D_B (n \cdot p)}{ n \cdot p} \\
& \times\sum_\helicity   \int \hd \omega \; \omega \; \hdelta_+(\omega)  a_\helicity (\omega n) e_\helicity ^{AB} + \textrm{h.c.}\,.
\]  
We proceed by invoking the equality
\[\label{eq:TTshear}
e_{\eta}^{AB} D_A D_B T(n) = 2 G e_{\eta}^{AB} \frac{ D_A (n \cdot p) D_B (n \cdot p)}{ n \cdot p}  \,,
\]
which relies on
$D_A D_B \v{n}= - \Omega_{AB} \v{n}$
and tracelessness of $e_{\eta}^{AB}$.
This allows us to expose the parameter $T(n)$ in the first integral in Eq.~\eqref{eq:SoftCharge1}.
For the $\omega$ integral, we define a shear operator (of dimension length) by  
\[\mathbb{f}_{AB}(u, \v{n}) = |\v{x}| \kappa  \, \hop_{\mu\nu}(x) \epsilon^\mu_A \epsilon^\nu_B \,,
\] 
with $\hop_{\mu\nu}$ given by the stationary phase approximation in the second Eq.~\eqref{eq:statPhase}. Then the soft charge becomes 
\[
Q_s = \frac{-1}{8 \pi G} \int \dd u \; \dd^2\Omega \; \dot{\mathbb{f}}^{AB}  \, D_A D_B T(n) \, . 
\]
As we will see below, this form of the soft charge operator makes 
the connection with the classical BMS charges more transparent.
%Expressing the soft charge operator in this form makes the 
%connection with the classical BMS charges more transparent as 
%we will see below. 
Before doing so, we rewrite the hard charge in a similar manner. 
We note that 
\[
\sum_\helicity  \int_0^\infty \hd \omega \, \omega^2 a^\dagger_\helicity (\omega n)  \, a_\helicity (\omega n)
= \frac{\pi}{4G} \int \dd u \; :\dot{\mathbb{f}}_{AB}\dot{\mathbb{f}}^{AB}: \,,
\]
using the normal-ordering symbol, and discarding terms involving $\int \dd \omega \,\theta(\omega)\theta(-\omega)$.
The hard charge is  
\[
Q_{h,0}  = - \frac{1}{32\pi G} \int \dd u \, \dd^2 \Omega \; : \dot{\mathbb{f}}_{AB}\dot{\mathbb{f}}^{AB}:  \, T(n) \ . 
\]
The total charge at null infinity, $Q_0 = Q_s + Q_{h,0}$, is then
\[
Q_0 =  \frac{-1}{8\pi G} \int  \dd u \,\dd^2\Omega \, T(n)  \left(D_A D_B \dot{\mathbb{f}}^{AB}  + \frac1{4} :\dot{\mathbb{f}}_{AB}\dot{\mathbb{f}}^{AB} : \right) \, .
\]
%%%%%%%%%%%%%%%%%%%%%%
\iffalse
\[
Q =  \frac{-1}{8\pi G} \int  \dd^2\Omega \, T(n) \int \dd u  \left(D_A D_B \dot{\mathbb{f}}^{AB}  + \frac1{4} :\dot{\mathbb{f}}_{AB}\dot{\mathbb{f}}^{AB} : \right) \, .
\]
\fi
%%%%%%%%%%%%%%%%%%%%%%%
The classical expectation value of this operator in the out-state $S\ket{\psi}$, $Q_\textrm{out} \equiv \braket{\psi | S^\dagger Q_0  S| \psi}$ is 
\[\label{eq:ChargeImpulse}
Q_\textrm{out} =
\frac{-1}{8\pi G}\int \dd u \, \dd^2\Omega \, T(n)
 \left(D_A D_B \dot{{f}}^{AB} + \frac14\dot{{f}}_{AB}\dot{{f}}^{AB}  \right) \,,
\]
%%%%%%%%%%%%%%%%%%%%%
\iffalse
\[\label{eq:ChargeImpulse}
Q_\textrm{out} =
\frac{-1}{8\pi G}\int \dd^2\Omega \, T(n)
\int \dd u  \left(D_A D_B \dot{{f}}^{AB} + \frac14\dot{{f}}_{AB}\dot{{f}}^{AB}  \right) \,,
\]
\fi
%%%%%%%%%%%%%%%%%%%%%
where the classical shear (in the future) is
\[\label{eq:OutShear}
f_{AB} \equiv \bra{\psi}S^\dagger {\mathbb{f}}_{AB} S \ket{\psi} 
\]
and we assumed that $S\ket{\psi}$ is a classical state, so that
\[
\bra{\psi}S^\dagger :\dot{\mathbb{f}}_{AB}\dot{\mathbb{f}}^{AB} : S \ket{\psi}=\bra{\psi}S^\dagger \dot{\mathbb{f}}_{AB} S \ket{\psi} \bra{\psi} S^\dagger \dot{\mathbb{f}}^{AB} S \ket{\psi}
\]
(see Ref.~\cite{Cristofoli:2021jas} for a discussion of these relations in terms of amplitudes). Notice that the forward scattering ($S=S^{\dagger}=1$) term in $f_{AB}$ does not vanish but instead contains the static $\mathcal{O}(G)$ contribution to the shear. However, the charge $Q_\textrm{out}$ only depends on $\dot{f}_{AB}$ so that it does not receive contributions from the forward scattering term. This allows us to identify
\[
Q_\textrm{out} = \Delta Q_0 \equiv \braket{\psi | S^\dagger Q_0  S| \psi} - \braket{\psi |  Q_0  | \psi},
\]
where we have subtracted the vanishing term $ \braket{\psi |  Q_0  | \psi}$.

To connect this to the classical BMS charges, consider the total variation of the classical supertranslation charge at null infinity
\[\label{eq:ClassicalDeltaQ}
\Delta Q_\textrm{cl}  \equiv \int \dd u \,\dot{Q}_\textrm{cl} = \int \dd u \, \hd^2 \Omega  \, \dot{m} \, T(n) \ . 
\]
Using the Bondi mass-loss formula which, in our conventions, is 
\[\label{ClassicalCharge}
\dot{m} = -\frac{1}{8 \pi G} \left( D_A D_B  \dot{f}^{AB} + \frac 14 \dot{f}_{AB}  \dot{f}^{AB} \right) \,,
\]
it is easy to see that $\Delta Q_\textrm{cl} $ coincides with Eq.\eqref{eq:ChargeImpulse}. We emphasise that this equality holds provided that there are no gravitons in the initial state $\ket{\psi}$. 

Note that the classical charge in Eq.~\eqref{eq:ClassicalDeltaQ} is defined by a surface integral at null infinity. 
In particular, it does not include contributions  $Q_{h,M}$ from massive particles located at future/past timelike infinity $i^{+}$ and $i^-$ respectively.
Such contributions are accounted for using the boundary-to-bulk propagators introduced in \cite{Campiglia:2015kxa,Campiglia:2015lxa,Campiglia:2015qka}.
In general BMS conservation laws require that the variation of the total BMS charge vanishes, i.e. 
\[
\Delta Q_0 + \Delta Q_{h, M} = 0,
\]
where $\Delta Q_{h, M}$ is the difference of the hard charges of the incoming and outgoing matter particles.

\section{Discussion}

The choice of BMS frame has observable consequences, as demonstrated by differences in physical predictions in the intrinsic and canonical frames.
Starting with the observation that low frequency mediators break assumptions built into the large-distance approximation for asymptotic observables, we defined new asymptotic states that resum their contribution. They turn out to be coherent states dressing the usual single-particle asymptotic states.
We showed that the argument $Q_s$ of the dressing phase operator, which is linear in the mediator's creation and annihilation operators, can be completed to an operator $Q=Q_s+Q_h$ that commutes with the scattering operator to all orders in the coupling, at least in the classical (large mass) limit. 

Using this conservation we found the exact contribution of the coherent state of low frequency mediators to the asymptotic radiation field, observing the close relation between the linearized Schwarzschild background and  
a particular BMS supertranslation in the far future. 
We strengthened this connection by explicitly showing that the final-state expectation value of the conserved operator $Q$ equals the (variation of the) classical BMS charge. This also makes contact with discussions of BMS symmetry of the $S$ matrix from a general relativity perspective, e.g.~\cite{Strominger:2013jfa, Campiglia:2015yka,He:2014laa,Ware:2013zja,Adamo:2014yya,Strominger:2017zoo}, and its connection to the Faddeev-Kulish formalism~\cite{Choi:2017bna, Choi:2017ylo} while effectively solving the supertranslation Ward identity for final-state observables. 

It would be interesting to understand how to include in the on-shell formalism more general BMS supertranslations than the Veneziano-Vilkovisky one, connecting the intrinsic and canonical BMS frames, perhaps by extending the expression for the soft charge to a more general function of angles (or of $p\cdot n$) and constructing the corresponding hard charge.
Similarly, it remains an interesting problem for the future to understand whether analogous results hold for superrotations \cite{deBoer:2003vf,Barnich:2009se,Barnich:2010ojg}. The corresponding Ward identities have been discussed in~\cite{Kapec:2014opa,Campiglia:2014yka} and their connection to classical soft theorems in~\cite{Laddha:2018myi,Laddha:2018vbn,Sahoo:2018lxl,Saha:2019tub,Sahoo:2020ryf}.

We have also evaluated the contribution of the low-frequency mediators to the orbital angular momentum radiated in a scattering process and recovered 
known results and extended them to all orders in the coupling.

While our main focus was on gravitationally-interacting 
massive particles, our analysis carries over, with minimal modifications, to other types of mediators. It would be interesting to explore the scalar charge model~\cite{Quinn:2000wa, Gralla:2021qaf}, in which one of the massive scalars interacts with gravity through a massless scalar. 
The asymptotic symmetries have been discussed in~\cite{Campiglia:2017dpg, Campiglia:2018see}.

One aim of our work was to include low frequency mediators which spoil the large $\omega |\bm x|$ assumption of the large-distance approximation. As mentioned, an alternative approach to this problem is to have the asymptotic states created and destroyed at finite distance as well as a finite IR cutoff for the mediator's frequency.
At finite, but not large distance (when $1/|\bm x|^2$ terms in the metric are relevant) BMS transformations are no longer expected to be symmetries of the metric~\cite{Flanagan:2015pxa}, and thus our analysis does not apply.   
Thus, exploring the relation between this and our approach should shed light on the interpretation of BMS transformations away from null infinity.

\section{Acknowledgements}

We thank Simon Caron-Huot, Hofie Hannesd\'ottir, Carlo Heissenberg, Aidan Herderschee, Anton Ilderton, Ricardo Monteiro and Fei Teng. 
DOC is supported by the STFC grant ``Particle Theory at the Higgs Centre''. 
RR~is supported by
the U.S.  Department of Energy (DOE) under award number~DE-SC00019066.
For the purpose of open access, the author has applied a Creative Commons Attribution (CC BY) licence to any Author Accepted Manuscript version arising from this submission.

\bibliography{letter}
\end{document}